\documentclass[12pt]{article}

\usepackage{cite}
\usepackage{epsfig}
\usepackage{amsmath}
\usepackage{tabularx}
\usepackage{pstricks}
\usepackage{array}

\def\mathswitch#1{\relax\ifmmode#1\else$#1$\fi}
\def\mathswitchr#1{\relax\ifmmode{\mathrm{#1}}\else$\mathrm{#1}$\fi}

\newcommand{\tev}{\,\, \mathrm{TeV}}
\newcommand{\gev}{\,\, \mathrm{GeV}}

\newcommand{\SLASH}[2]{\makebox[#2ex][l]{$#1$}/}

\newcommand{\pslash}{\SLASH{p}{.2}}

\newcommand{\Eslash}{\SLASH{E}{.3}\,}

\newcommand{\anc}{\rule{0mm}{0mm}}
\newcommand{\lesim}{\,\raisebox{-.1ex}{$_{\textstyle<}\atop^{\textstyle\sim}$}\,}
\newcommand{\gesim}{\,{_{\textstyle
>}\atop^{\textstyle\sim}}\,}

\newcommand{\mycaption}[1]{\caption{\sl #1}}

\begin{document}
\thispagestyle{empty}

\def\thefootnote{\fnsymbol{footnote}}

\begin{flushright}
PITT-PACC-1208 \\ 
CETUP*-12/001 \,
\end{flushright}

\vspace{1cm}

\begin{center}

{\Large\sc {\bf New Physics from the Top at the LHC}} 

\vspace*{1.5cm}

Chien-Yi~Chen$^1$, Ayres~Freitas$^2$, Tao~Han$^2$, and Keith~S.~M.~Lee$^2$

\vspace*{1cm}

{\sl $^1$ Department of Physics, Carnegie Mellon University, Pittsburgh, PA
%15213, USA
\\[1em]
\sl $^2$
PITTsburgh Particle physics, Astrophysics, and Cosmology Center (PITT PACC),\\
Department of Physics \& Astronomy, University of Pittsburgh,
Pittsburgh, PA %15260
%3941 O'Hara St, Pittsburgh, PA 15260, USA
}

\end{center}

\vspace*{2.5cm}

\begin{abstract}

The top quark may hold the key to new physics associated with the electroweak
symmetry-breaking sector, given its large mass and enhanced coupling to the
Higgs sector. 
We systematically categorize generic interactions of a new particle that
couples to the top quark and a neutral particle, which is assumed to be
heavy and stable, thus serving as a candidate for cold dark matter. 
The experimental signatures for new physics involving top quarks and its
partners at the Large Hadron Collider (LHC) may be distinctive, yet challenging to disentangle. 
We optimize the search strategy at the LHC for the decay of the new particle 
to a top quark plus missing energy and propose the study of its properties, 
such as its spin and couplings. We find that, at 14~TeV with an 
integrated luminosity of 100 fb$^{-1}$, a spin-zero top partner can be 
observed at the 5$\sigma$ level for a mass of 675~GeV. A spin-zero particle
can be differentiated from spin-1/2 and spin-1 particles at the 5$\sigma$ 
level with a luminosity of 10~fb$^{-1}$.

\end{abstract}

\setcounter{page}{0}
\setcounter{footnote}{0}

\newpage

%%%%%%%%%%%%%%%%%%%%%%%%%%%%%%%%%%%%%%%%%%%%%%%%%%%%%%%%%%%%%%%%%%%%%%%%%%%%%%

\section{Introduction}

The top quark may be a window to physics beyond the Standard Model (SM). Its 
mass near the electroweak scale and its large coupling to the Higgs boson may 
be crucial to understanding the electroweak sector beyond the SM. 
Now that the SM-like Higgs boson has been observed at the Large Hadron Collider
(LHC) \cite{Higgs} with a relatively light mass of about 125 GeV, 
the assumed ``naturalness'' of the Higgs sector \cite{Giudice:2008bi} suggests 
the existence of a partner of the top quark below or near the TeV scale, motivating theories such as  weak-scale 
supersymmetry, Little Higgs, and extra dimensions (either warped or universal). Vacuum stability of the electroweak potential also indicates the need for new physics to balance the large top-quark contribution. 
The top quark hence provides a possible early indicator of new physics and a 
good probe of a wide variety of new-physics scenarios.  

The LHC is a top factory, producing a hundred times more $t \bar{t}$ pairs 
from QCD processes than were produced at the Tevatron. Top-quark production 
is well understood in the SM. Thus any new physics contributions will be 
on top of a well-known and well-measured, albeit large, background. 
With the discovery era ushered in by the LHC, it would be prudent to keep the 
initial search as general as possible.

In this work, we take a model-independent approach to searching for new physics
processes of the form $$pp\to Y\bar{Y} \to t\bar{t}XX,$$ where $Y$ is a massive
new particle with the same gauge quantum numbers as the top quark and $X$ is an
electrically and color neutral stable particle. The weakly interacting
$X$ could be a constituent of dark matter, which would manifest itself as
missing energy in a collider detector.
We systematically consider different spin configurations (0,
1/2, and 1) for the new particles $Y$ and $X$. Each combination is exemplified
by particles in well-motivated new-physics models (see the next section for
details). For example, in the Minimal Supersymmetric Standard Model (MSSM) 
$Y$ could be a scalar top and $X$ the lightest neutralino. This case has 
been studied extensively in the literature
(see, for example, 
Refs.~\cite{Meade:2006dw,Han:2008gy,stoptag1,stopsearch2,stoptag2,stopsearch3,stophad2}). 
However, we do not limit ourselves to specific particles in a particular model;
rather, we undertake a general categorization, assuming merely a mass
accessible at the LHC and a discrete symmetry that ensures the stability of $X$. 
For simplicity, we restrict consideration to processes that involve only the top
partner, $Y$, and the dark-matter candidate, $X$, as new particles. 

In order to distinguish experimentally between the different possibilities, one
needs to determine the spins and couplings of the new particles $Y$ and $X$.  
In this paper, several observables for this purpose are proposed and their
usefulness is demonstrated in a realistic Monte Carlo simulation. To avoid
ambiguities due to model-dependent branching fractions, we do not consider
the total cross section in this set of variables.

The paper is organized as follows. In section~\ref{setup} we introduce the
model-independent classification of new-physics top partners and their
interactions. The production of these particles at the LHC is discussed in
section~\ref{channels}, while the current bounds from collider searches are
summarized in section~\ref{bounds}. In section~\ref{sigsel}, the expected reach
of the LHC for this class of processes is analyzed through a detailed 
Monte Carlo simulation. The determination of relevant properties of the 
new particles, such as mass, spin and couplings, and the discrimination 
between models are discussed in section \ref{props}. 
Finally, conclusions are presented in section~\ref{concl}.

%%%%%%%%%%%%%%%%%%%%%%%%%%%%%%%%%%%%%%%%%%%%%%%%%%%%%%%%%%%%%%%%%%%%%%%%%%%%%%

\section{New Particles and their Couplings to the Top}
\label{setup}

\begin{figure}[tb]
\centering
\psfig{figure=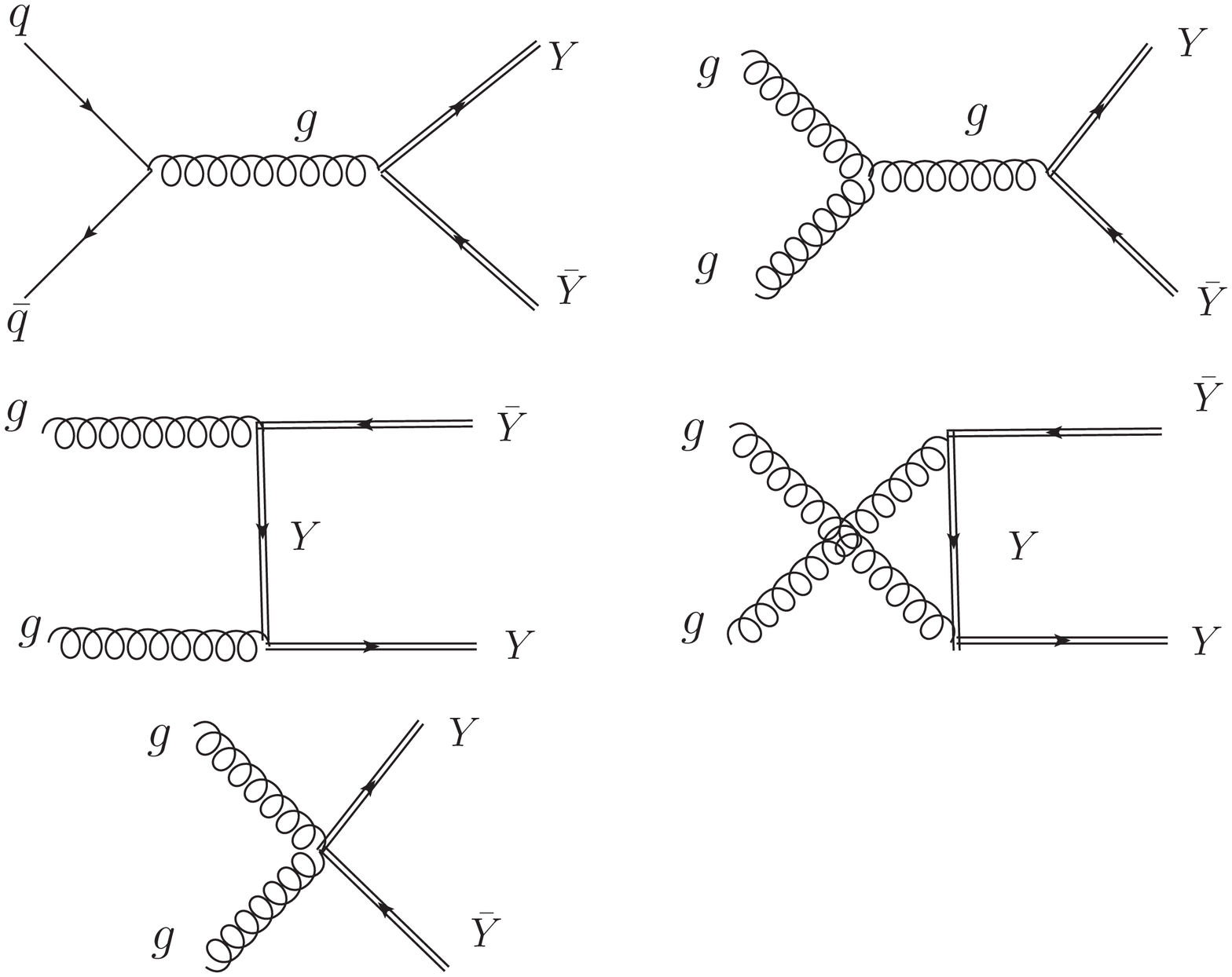, width=11.4cm}
\rule{.5pt}{8.5cm}\hfill
\raisebox{2.3cm}{\psfig{figure=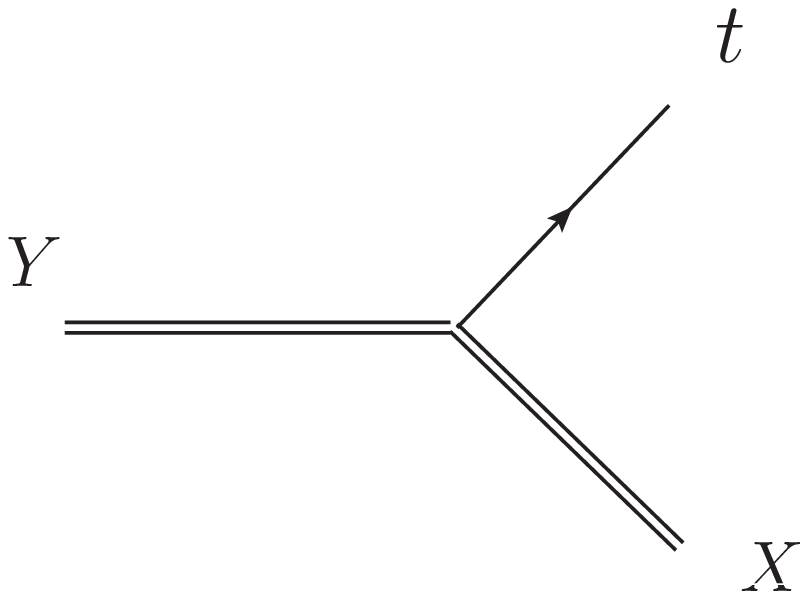, width=4.6cm, bb=163 542 400 721}}
\mycaption{Diagrams corresponding to the pair production (left) and the 
decay (right) of the color triplet $Y$.
Double lines denote new particles, while single lines 
denote SM particles.}
\label{fig:prodxy}
\end{figure}

Colored particles can be copiously produced at the LHC by strong QCD 
interactions. 
Let $Y$ denote a new color-triplet particle with charge +2/3. 
$Y$ and its antiparticle can be produced at leading order in QCD by the 
processes shown in Fig.~\ref{fig:prodxy}~(left). We shall not consider the
production of a single new particle via Yukawa-type interactions: since they 
are strongly model-dependent and are subject to strong constraints from 
flavor physics, it is assumed that such vertices are forbidden by a
discrete symmetry. 
$Y$ decays to a new particle that is a color singlet, denoted $X$ [see
Fig.~\ref{fig:prodxy}~(right)], which will show as missing energy in a collider
experiment.

%-----------------------------------------------------------------------------
%-----------------------------------------------------------------------------

There are four possible combinations of spins that allow a coupling between 
$X$, $Y$ and the SM top quark, $t$. These are listed, with the 
relevant couplings and sample model decays, in Table~\ref{tab:models1}. 
For fermions we allow a general chirality structure. We shall henceforth 
refer to these scenarios as models i, ii, iii, and iv.

%-----------------------------------------------------------------------------
\begin{table}[t]
\renewcommand{\arraystretch}{1.4}
\centering
\begin{tabular}{|l||c|c||c|c||ll|}
\hline
 & $Y$ & $X$ & $GYY$ & $XYt$ & 
 \multicolumn{2}{l|}{sample model and decay} \\[-1ex]
 & $J_{Y},\ I_{\rm SU(3)}$ & $J_{X},\ I_{\rm SU(3)}$ &
 coupling & coupling & \multicolumn{2}{l|}{$Y \to t X$} \\
\hline\hline
i & 0, {\bf 3} & $\frac{1}{2}$, {\bf 1} & 
 $G^{a\mu} Y^* \!\overleftrightarrow{\partial}_{\!\!\!\mu} T^a Y $ &
 $\overline{X} \Gamma t \, Y^*$ &
 MSSM & $\tilde{t} \to t \tilde{\chi}_1^0$ \\
\hline
ii & $\frac{1}{2}$, {\bf 3} & 0, {\bf 1} & 
 $\overline{Y} \SLASH{G}{.3}\,^a T^a Y $ &
 $\overline{Y} \Gamma t \, X$ &
 UED & $t_{\rm KK} \to t \gamma_{H,\rm KK}$ \\
\hline
iii & $\frac{1}{2}$, {\bf 3} & 1, {\bf 1} & 
 $\overline{Y} \SLASH{G}{.3}\,^a T^a Y $ &
 $\overline{Y}\SLASH{X}{.4}\; \Gamma t$ &
 UED & $t_{\rm KK} \to t \gamma_{\rm KK}$ \\
\hline
iv & 1, {\bf 3} & $\frac{1}{2}$, {\bf 1} &
 $S_3[G,Y,Y^*]$ &
$\overline{X} \SLASH{Y}{.2}\,^* \Gamma t$ &
 \cite{Cai:2008ss} & $\vec{Q} \to t \tilde{\chi}^0_1$ \\
\hline
\end{tabular}

\vspace{1ex}
\begin{tabular}{l}
$ 
\Gamma \equiv a_L P_L + a_R P_R \,,\quad
A \!\overleftrightarrow{\partial}_{\!\!\!\mu} B \equiv
A (\partial_\mu B) - (\partial_\mu A) B
$\\
$
S_3[G,Y,Y^*] \equiv T^{a} \left[
  G_\mu^a \, Y^{*}_\nu \!\overleftrightarrow{\partial}^{\!\!\!\mu} Y^{\nu} +
  G_\mu^a \, Y^{\mu*} \!\overleftarrow{\partial}^{\!\!\!\nu} Y_{\nu} -
  G_\mu^a \, Y^{*}_\nu \!\overrightarrow{\partial}^{\!\!\!\nu} Y^{\mu}
\right]
$
\end{tabular}
\vspace{-1ex}
\mycaption{Quantum numbers and couplings of the new particles $X$ and $Y$, 
which interact with the SM top quark, $t$. 
In the last column, $\tilde{t}$ and $\tilde{\chi}^0_1$ are the scalar top and
lightest neutralino in the MSSM, respectively \cite{susy}. $t_{\rm KK}$, 
$\gamma_{\rm KK}$, and $\gamma_{H,\rm KK}$ are the first-level Kaluza-Klein 
excitations of the top, the photon, and an extra-dimensional component of 
a photon, respectively, in universal extra dimensions (UED) \cite{ued}. 
Finally, $\vec{Q}$ is the vector superpartner in a supersymmetric model with 
an extended gauge sector \cite{Cai:2008ss}.}
\label{tab:models1}
\end{table}
%-----------------------------------------------------------------------------

Let us elaborate on the unusual case in which $Y$ is a vector color
triplet, possibly arising as a bound state from strong dynamics or from a
special kind of supersymmetric model \cite{Cai:2008ss}.
The kinetic term is 
\begin{align}
{\cal L}_\mathrm{kin}  = - \frac{1}{2} (F_{\mu\nu})^\dagger F^{\mu\nu}
\,, \quad \ 
F_{\mu\nu}  = D_\mu Y_\nu - D_\nu Y_\mu \,,
\end{align}
where 
$D_\mu= \partial_\mu - i g T_a G_\mu^a$.
Then the $Y$-$Y$-gluon interaction term is
\begin{align}
{\cal L}_{YYG} & = \frac{1}{2}i g (T_a)_{ji} \left( 
(\partial^\mu \bar{Y}^\nu_j - \partial^\nu\bar{Y}^\mu_j) 
(G_\mu^a Y_{\nu i} - G_\nu^a Y_{\mu i})
- \mathrm{h.c.} \right)
\,.
\end{align}
The resulting Feynman rule is 
\begin{align}
\begin{array}{l} \includegraphics[width=2in]{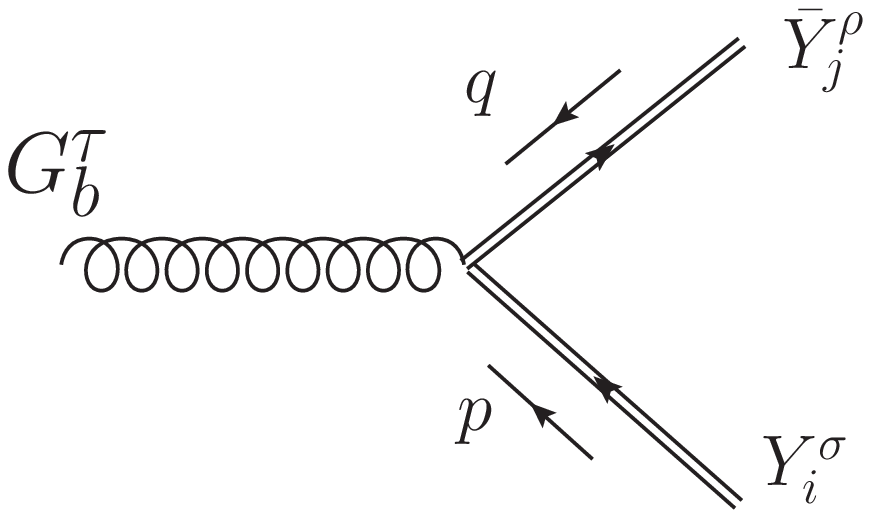} 
\end{array} 
& = i g (T_b)_{ji}\left((q-p)^\tau g^{\sigma\rho} + p^\rho g^{\sigma\tau}
    - q^\sigma g^{\rho\tau}\right)
\,.
\end{align}
Likewise, the $Y$-$Y$-gluon-gluon interaction term is
\begin{align}
{\cal L}_{YYGG} & = -\frac{g^2}{2} 
(G^\mu_b \bar{Y}^\nu - G^\nu_b \bar{Y}^\mu)
T_b T_a (G_\mu^a Y_\nu - G_\nu^a Y_\mu)
\,.
\end{align}
The resulting Feynman rule is 
\begin{align}
\begin{array}{l} \includegraphics[width=1.8in]{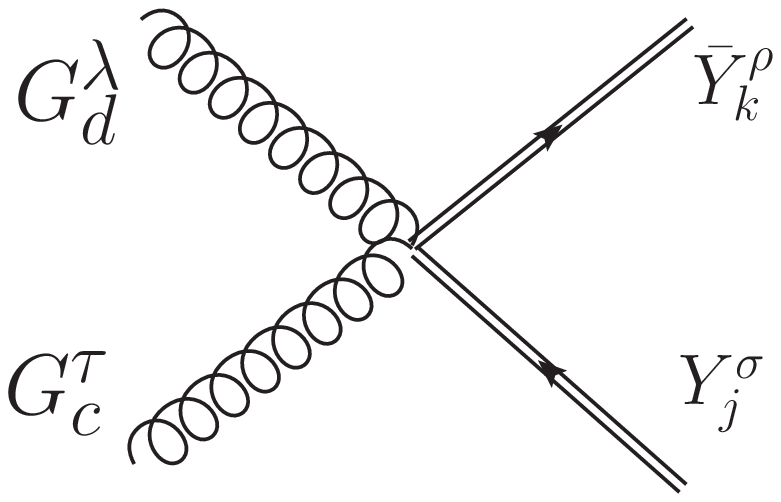} 
\end{array} 
& =  -i g^2 \left((T_c T_d + T_d T_c)_{kj} g^{\tau\lambda}g^{\rho\sigma}
- (T_c T_d)_{kj} g^{\tau\sigma}g^{\lambda\rho} 
- (T_d T_c)_{kj} g^{\tau\rho}g^{\lambda\sigma} \right)
\,.
\end{align}

%%%%%%%%%%%%%%%%%%%%%%%%%%%%%%%%%%%%%%%%%%%%%%%%%%%%%%%%%%%%%%%%%%%%%%%%%%%%%%

\section{Color-Triplet Top-Partner Production}
\label{channels}

%-----------------------------------------------------------------------------
\begin{figure}
\anc\hspace{10mm}\makebox[8.5cm][l]{\small $\sqrt{s}=8$~TeV}%
\makebox[6.5cm][l]{\small $\sqrt{s}=14$~TeV}\\[.3ex]
\centering
\epsfig{figure=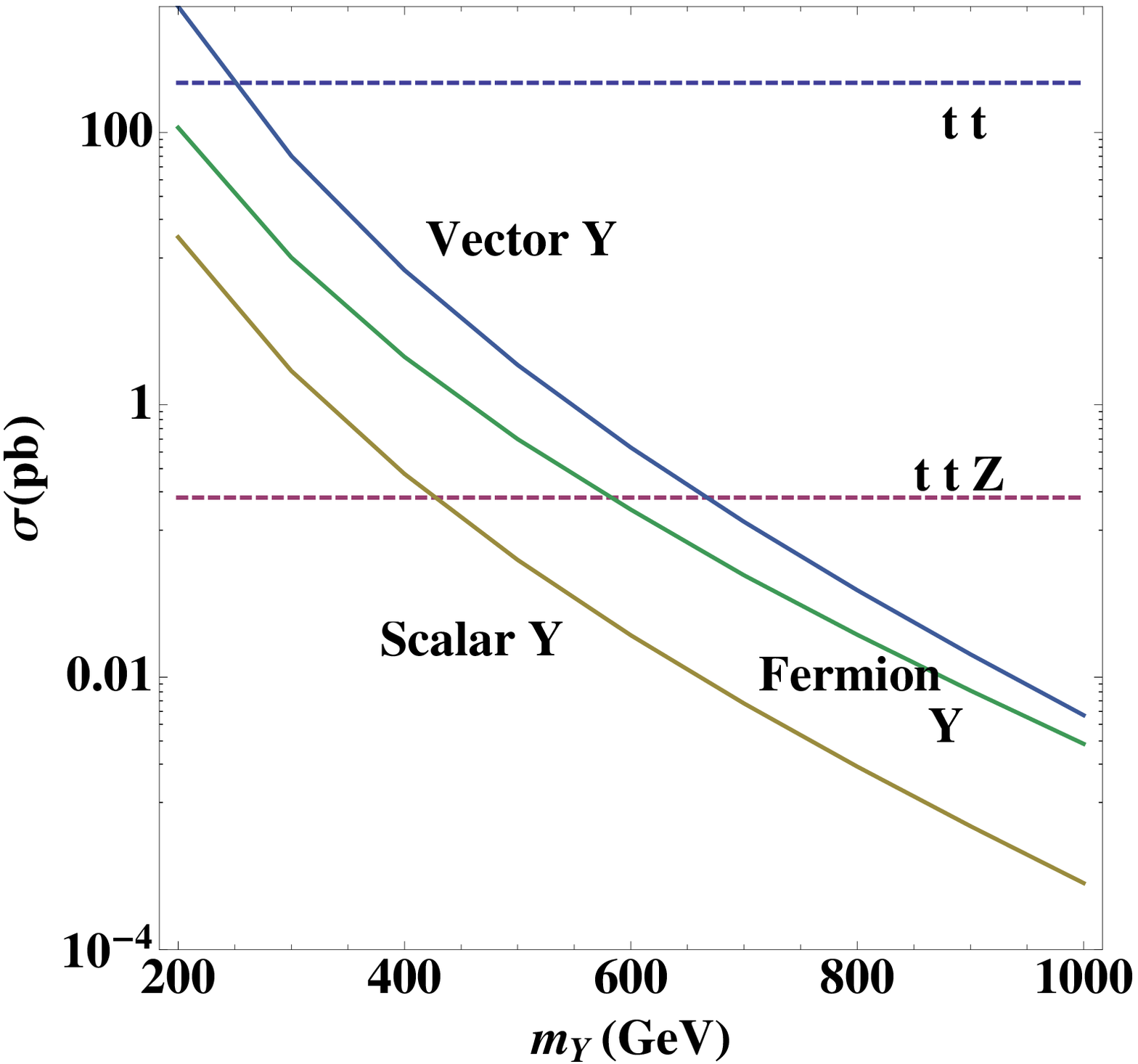, height=7.4cm}
\hfill
\epsfig{figure=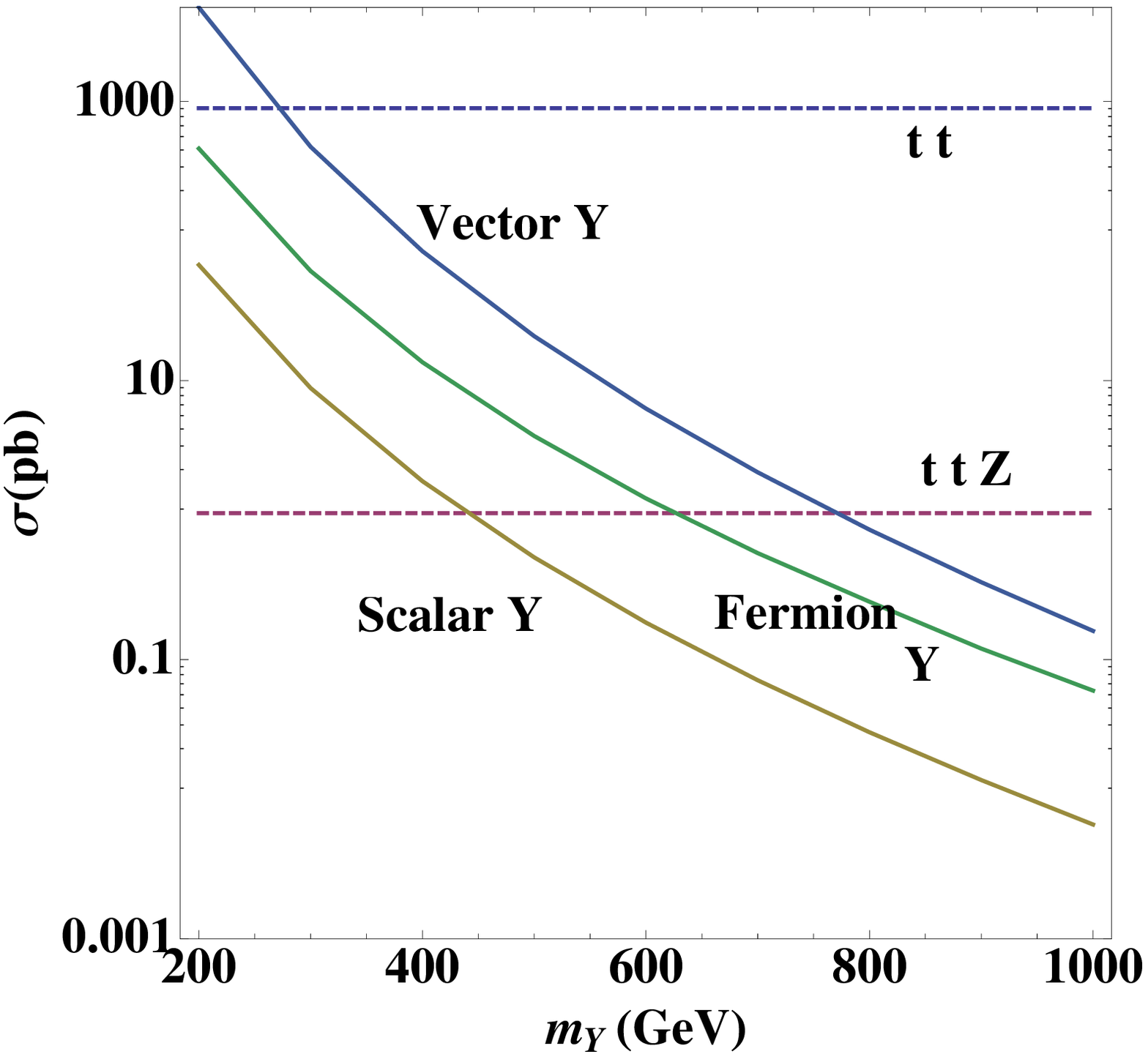, height=7.4cm}
\vspace{-.3em}
\mycaption{Production cross sections for $pp\to Y\bar{Y}$ at the LHC for
8~TeV (left) and 14~TeV (right), as a function of the mass $m_Y$, 
for different $Y$ spins. The leading SM backgrounds are indicated by 
horizontal lines. 
\label{plot:prody}}
\end{figure}
%-----------------------------------------------------------------------------

The dominant modes for production of the top partner in hadronic collisions 
are the QCD subprocesses
\begin{align}
 q \bar{q} \,, gg & \rightarrow  Y \bar{Y} \,.
\end{align}
We restrict ourselves to the first- and second-generation quarks $q = u,d,c,s$
and use the CTEQ~6L1 parton distribution functions (PDFs) \cite{Pumplin:2002vw}, with the
factorization scale set to $m_Y$. %$\sqrt{s}/2 = m_Y$.
For $m_Y \sim 200$--1000~GeV, the dominant subprocess is $gg \to Y \bar{Y}$,
which is about one order of magnitude larger than $u \bar{u},d \bar{d} \to Y
\bar{Y}$. The channels $c \bar{c},s \bar{s} \to Y
\bar{Y}$ are suppressed by roughly one additional order of magnitude.

The total QCD production cross section at the LHC as a function of the mass of 
the $Y$ is shown in Fig.~\ref{plot:prody}, for the cases in which $Y$ has
spin $0$, $1/2$, and $1$. The plots include next-to-leading order (NLO) and
resummed next-to-leading logarithmic (NLL) QCD corrections for the scalar $Y$
\cite{scalxsec} and NLO and NNLL corrections for the
fermionic $Y$ \cite{fermxsec}. The QCD corrections for vector $Y$ production 
have not yet been calculated; we use the $K$-factor for the scalar $Y$ 
(1.77 at $\sqrt{s}=8\tev$, 1.56 at $\sqrt{s}=14\tev$), since
the two cases share similar leading-order Feynman diagrams.

The cross section for the fermion is about an order of magnitude larger than 
that of the scalar, because of the fermion's extra spin degrees of freedom 
and threshold effects. In the $s$-channel, the scalar is produced as a 
$p$-wave with a velocity dependence of $\sigma \sim  \beta^3$, 
whereas the fermion is produced as an $s$-wave with $\sigma \sim \beta$. 
Thus, the ratio of the cross sections of the fermion 
and scalar is larger at small values of $\beta$, that is, when the mass
of the $Y$ is large. This relative enhancement of the fermionic $Y\bar{Y}$
production is particularly pronounced when the $Y$ particles are produced 
mostly near the threshold limit.
Note that, although the curves for the vector and scalar appear to be 
parallel on the logarithmic scale, their ratio varies from about
34 to 16 in the mass range shown.

%%%%%%%%%%%%%%%%%%%%%%%%%%%%%%%%%%%%%%%%%%%%%%%%%%%%%%%%%%%%%%%%%%%%%%%%%%%%%%

\section{Current Bounds from the Tevatron and LHC}
\label{bounds}

As Fig.~\ref{fig:prodxy} indicates, the top-quark partner, $Y$, decays to a 
top quark plus a neutral particle, $X$. The discrete symmetry implies that
$X$ is stable and leads to missing-energy events. Thus, the 
signal is $t\bar t$ plus missing energy. Searches for supersymmetric scalar tops 
at the Tevatron \cite{cdf-tt,Abazov:2010wq} and the LHC
\cite{atlas-tt,atlas-tt2,cms-tt2} put 
constraints on the allowed parameter space for the class of processes 
considered here. Additional, though generally weaker, bounds also arise
from general searches for signals with jets and missing energy \cite{lhcsusy2}.

Currently, the strongest constraints arise from scalar top searches at ATLAS
using 4.7~fb$^{-1}$ of data taken at $\sqrt{s}=7\tev$ \cite{atlas-tt2}. For
$m_Y \gg m_X$, they put a lower bound $m_Y \gesim 500\gev$ on a
scalar $Y$. By taking into account the different production cross sections 
for scalars, fermions, and vectors (see Fig.~\ref{plot:prody}), one can 
translate the results of Refs.~\cite{atlas-tt2} into a limit of 
$m_Y \gesim 650\gev$ for a fermionic top partner and $m_Y \gesim 730\gev$ 
for a vector top.

The bounds are summarized in Table \ref{tab:bounds}.
It should be pointed out that the limits for fermionic and vector $Y$ 
are simple estimates from theoretical considerations. For more robust results, 
a detailed experimental analysis of these scenarios needs to be performed.
%-----------------------------------------------------------------------------
\begin{table}[t]
\renewcommand{\arraystretch}{1.3}
\centering
\begin{tabular}{|c||c|}
\hline
 $J_Y$ & Limit on $m_Y$ \\
\hline
 0 & $\gesim 500\gev$ \\
 1/2 & $\gesim 650\gev$ \\
 1 & $\gesim 730\gev$ \\
\hline
\end{tabular}
\mycaption{Experimental bounds on the mass of particle $Y$ for different
spins, $J_Y$, under the assumption $m_Y \gg m_X$. These estimates are based on 
the ATLAS results from Refs.~\cite{atlas-tt2}.
}
\label{tab:bounds}
\end{table}
%-----------------------------------------------------------------------------

For larger values of $m_X$, that is, smaller mass differences $m_Y-m_X$,
the limits become weaker. The excluded region
in the $m_Y$--$m_X$ mass plane for scalar $Y$ particles will be 
shown in  the next section (see Fig.~\ref{fig:sb}).

%%%%%%%%%%%%%%%%%%%%%%%%%%%%%%%%%%%%%%%%%%%%%%%%%%%%%%%%%%%%%%%%%%%%%%%%%%%%%%

\section{Signal Observability at the LHC}
\label{sigsel}

As the previous section discusses, we consider new physics signals of
the type $t\bar{t}+\Eslash$. For the leading channel, 
in which the top quarks decay hadronically \cite{Meade:2006dw},  
the signal receives large 
backgrounds from SM processes with multiple QCD jets. To suppress QCD 
backgrounds, we consider the semileptonic channel \cite{Han:2008gy}, in which one of the tops 
decays hadronically and the other decays leptonically, namely,
\begin{align}
pp &\to Y \bar{Y} \to t X \  \bar{t} X \to b j_1 j_2 \ \bar{b} \ell^- \bar{\nu}_\ell \ XX + \mathrm{h.c.} \,
\qquad (\ell = e,\mu).
\end{align}
This channel is beneficial because of its sizeable branching fraction
and the identification of both $t$ and $\bar t$. The dominant background 
processes are 
\begin{equation}
t\bar{t}, \quad t\bar{t}Z\ {\rm (with }\ Z\to\nu\bar{\nu} ),\quad {\rm and}\quad Wb\bar{b}jj\ {\rm (with}\ W\to \ell\nu_\ell). 
\end{equation}
The cross sections for the first two backgrounds (without branching fractions)
are shown in Fig.~\ref{plot:prody} as horizontal lines, including NLO 
corrections for $t\bar{t}Z$ \cite{ttz} and NLO+NNLL effects for 
$t\bar{t}$ \cite{tt}.

The separation of signal and background in the semileptonic channel has been
studied previously in the literature
\cite{Han:2008gy,stoptag1,cdf-tt,atlas-tt,atlas-tt2,cms-tt2}.
Here, we reanalyze the signal selection with the purpose of developing 
optimized selection cuts in a phenomenologically realistic
simulation setup. Our signal selection follows the strategy of
Ref.~\cite{Han:2008gy}, but we include QCD parton showering and detector smearing effects.
As a result, we find that we need to adjust the choice of cuts to account for 
the effect of QCD radiation\footnote{Very recently, several papers have 
appeared that pursue a similar goal in the context of the MSSM, using 
traditional selection cuts \cite{stopsearch2,stopsearch3} and top-jet tagging 
techniques \cite{stoptag2}. Our results for the signal observability are 
comparable to Refs.~\cite{stopsearch2,stoptag2}, but significantly better than
Ref.~\cite{stopsearch3}.}.

Jets have been clustered via a cone algorithm with cone size 0.4. To simulate
detector resolution effects, we have smeared the jet energy with a Gaussian
distribution of width $0.5\times\sqrt{E}$, where $E$ is the jet energy in units of GeV.
A $b$-tagging efficiency of 70\% \cite{btag} has been assumed.
We have applied the following set of cuts to identify the signal signature and 
reduce the SM backgrounds.
\begin{equation}
\text{\it Cut(1):} \hspace{3em}
\begin{aligned}[c]
&\text{exactly one lepton $\ell=e,\mu$ with } E_T^\ell > 20\gev, \; |\eta_\ell| < 2.5; \\
&\text{at least two light jets with } E_T^j > 25\gev, \; |\eta_j| < 2.5; \\
&\text{exactly two b-tagged jets with } E_T^b > 30\gev, \; |\eta_b| < 2.5; \\
&\Delta R_{jj}, \Delta R_{bj}, \Delta R_{bb} > 0.4, \qquad
 \Delta R_{\ell j} = \Delta R_{\ell b} = 0.3; \\
&70\gev < m_{jj} < 90\gev, \quad 120\gev < m_t^{\rm had} < 180\gev; \\
&\Eslash > 25\gev. 
\end{aligned} \hspace{3em}
\label{eq:cut1}
\end{equation}
Here $b$ and $j$ stand for a jet with or without a $b$-tag, and $E_T^i$ and 
$\eta_i$ are the transverse energy and pseudorapidity of object $i$. 
$\Delta R = \sqrt{(\Delta \eta)^2 + (\Delta \phi)^2}$ describes the angular 
separation between two jets. $m_t^{\rm had}$ is computed from either
the $bjj$ or the $\bar{b}jj$ invariant mass, namely, whichever yields the 
value closer to the true top-quark mass, $m_t$, in a given event. 
$\Eslash$ is the missing transverse energy.

Events for the partonic signal process and $t\bar{t}Z$ background have been
generated with {\sc CalcHEP 3.2.5} \cite{calchep} and passed to 
{\sc Pythia 6.4} \cite{pythia} for parton showering and jet clustering. 
The $t\bar{t}$ background has been simulated with {\sc Pythia}. It was shown in 
Ref.~\cite{Han:2008gy} that the $Wb\bar{b}jj$ background can be reduced 
effectively with invariant-mass cuts on the $jj$ for a $W$ selection and 
$bjj$ for a top-quark selection. 
We have thus neglected this process in our simulation.

With the set of cuts in \eqref{eq:cut1}, which we shall refer to as 
\emph{Cut(1)}, a good signal-to-background ratio is achieved
for small values of $m_Y$, when the $Y\bar{Y}$ production cross section is 
large. %We shall refer to this set of cuts as \emph{Cut(1)}.
For larger values of $m_Y$, additional cuts are required to suppress the SM 
background sufficiently. It turns out that the following two variables are 
useful for this purpose: the missing transverse energy, $\Eslash$, and
the transverse mass of the lepton--missing-momentum system,
\begin{align}
M_T^{\ell,\rm miss} \equiv \sqrt{(E_{\ell T} + \Eslash)^2 - 
 (\mbox{\bf p}_{\ell T} + \SLASH{\mbox{\bf p}}{.2}_{\,T})^2}.
\label{eq:mtw}
\end{align}
The optimal cut values depend on the collider energy:

%-----------------------------------------------------------------------------
\begin{figure}[tb]
\centering
\psfig{figure=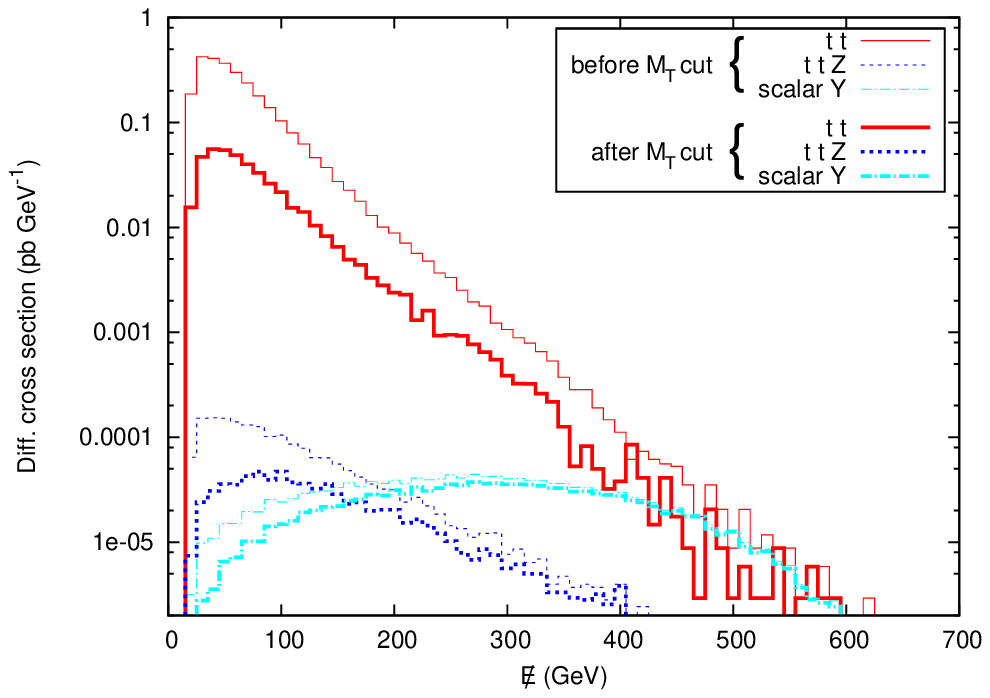, width=12cm}
\vspace{-.5em}
\mycaption{Differential cross sections for the SM backgrounds $t
\bar{t}$ (red, solid) and $t \bar{t} Z$ (blue, dashed) and the signal scalar Y
with $(M_Y,M_X) = (600,10)\gev$ (cyan, dash-dotted) before and after the cut
$M_T^{\ell,\rm miss} > 90 \gev$. Distributions after the $M_T^{\ell,\rm miss}$
cut are shown in bold lines.
\label{fig:METcomb}}
\end{figure}
%-----------------------------------------------------------------------------

\begin{itemize}

\item For $\sqrt{s}=14\tev$, the choice
\begin{align}
\text{\it Cut(2h)} \quad&=\quad \text{\it Cut(1)} \quad\text{plus}\quad 
\Eslash > 350\gev \quad\text{and}\quad M_T^{\ell,\rm miss} > 90\gev \nonumber
\end{align}
has been found to be effective for $m_Y \sim 600\gev$. It can be understood as
follows. The cut $M_T^{\ell,\rm miss} >90$ GeV is necessary because a large
amount of missing energy in the SM backgrounds corresponds to neutrinos from the
leptonic decay of the W boson. From Fig.~\ref{fig:METcomb} one can see that the
$M_T^{\ell,\rm miss}$ cut reduces the SM backgrounds dramatically, especially in
the low-$\Eslash$ region. However, the signal events remain virtually the same
after this cut: only those in the low-$\Eslash$ region are slightly affected.
Moreover, for the signal there is a plateau between 200 and 400~GeV in Fig~\ref
{fig:METcomb}. The cut $\Eslash >$ 350~GeV has been chosen because above 
350~GeV the backgrounds are suppressed considerably. In practice, we have
applied either \emph{Cut(1)} or \emph{Cut(2h)}, whichever produces the larger
statistical significance $S/\sqrt{B}$ for a given parameter point $(m_Y,\,m_X)$.
Here, $S$ and $B$ denote the number of signal and background events after cuts.

\item For $\sqrt{s}=8\tev$, we have used either \emph{Cut(1)}, or 
\begin{align} 
\text{\it Cut(2l)} \quad&=\quad \text{\it Cut(1)} \quad\text{plus}\quad 
\Eslash > 200\gev \quad\text{and}\quad M_T^{\ell,\rm miss} > 145\gev, \nonumber
\intertext{or}
\text{\it Cut(2l')} \quad&=\quad \text{\it Cut(1)} \quad\text{plus}\quad \Eslash > 300\gev
\quad\text{and}\quad M_T^{\ell,\rm miss} > 185\gev, \nonumber
\end{align}
whichever results in the largest significance. \emph{Cut(2l)} and 
\emph{Cut(2l')} have been optimized for $m_Y \sim 400\gev$ and 
$m_Y\sim 500\gev$, respectively.

Because the signal cross section is lower for $\sqrt{s}=8\tev$ than for
$\sqrt{s}=14\tev$, we have lowered the $\Eslash$ cut to ensure that
a sufficient number of signal events passes. However, the looser $\Eslash$
cut also results in a larger background event yield, so that it is advantageous
to apply a stronger cut on $M_T^{\ell,\rm miss}$ to improve the signal
significance.

\end{itemize}
%
%-----------------------------------------------------------------------------
\begin{figure}[tb]
\centering
\epsfig{figure=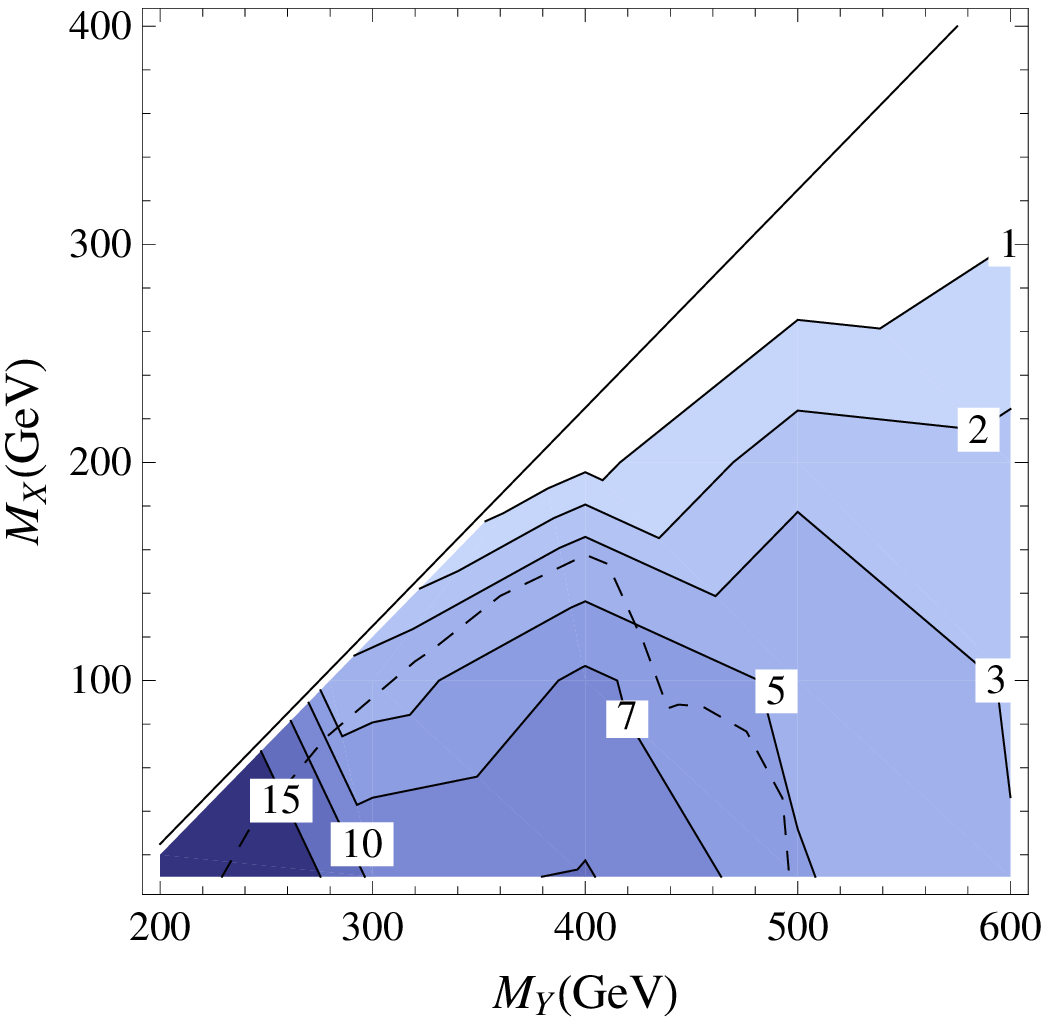, width=8cm}
\hfill
\epsfig{figure=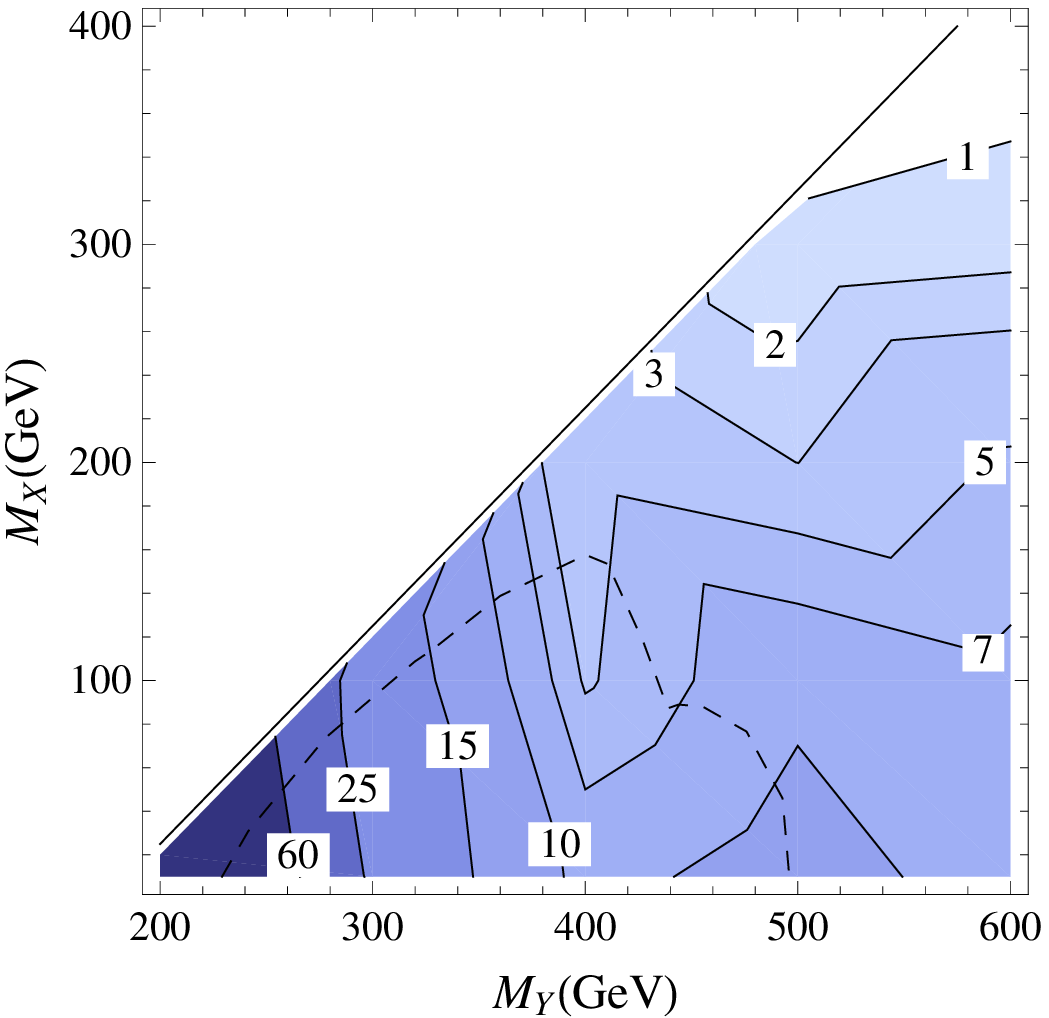, width=8cm}%
\vspace{-.5em}
\mycaption{Expected statistical significance for model i, as a function of the
masses of $X$ and $Y$. The left panel corresponds to $\sqrt{s}=8\tev$ and
${\cal L}=20$ fb$^{-1}$, while the right panel corresponds to 
$\sqrt{s}=14\tev$ and ${\cal L}=100$ fb$^{-1}$. The dashed line shows the
current exclusion limit at the 95\% confidence level from 
Refs.~\cite{atlas-tt2}.}
\label{fig:sb}
\end{figure}
%-----------------------------------------------------------------------------
Figure~\ref{fig:sb} shows the statistical significance that
can be achieved with these cuts for model i (scalar $Y$ and fermionic $X$), 
for different values of $m_Y$ and $m_X$. The significance is determined
according to $S/\sqrt{B}$ if $B>10$, whereas Poisson
statistics is used for very low event yields ($B\lesim 10$).
For the other model combinations, ii--iv, the statistical
significance can be obtained by scaling the values in Fig.~\ref{fig:sb} with 
the production cross sections in Fig.~\ref{plot:prody}.

As the figure shows, the statistical significance is relatively large in 
the following two regions of the mass plane:
\begin{enumerate}
\item Small values of $m_Y$, in which case the signal selection efficiency 
is almost independent of $m_X$. Here, the $Y\bar{Y}$ pair, which recoils 
against an initial-state jet, is typically produced with a sizeable boost. 
This boost leads to a fairly large missing momentum, which helps to 
discriminate the signal from the $t\bar{t}$ background. Our signal selection
works even near the kinematic threshold, where $m_Y-m_X$ approaches $m_t$, in
contrast to the experimental analysis by ATLAS \cite{atlas-tt2}, where such
events have been cut away with a strong cut on the azimuthal angle between
the leading two jets and the missing momentum\footnote{Note, however,
that we generated Fig.~\ref{fig:sb} by running dedicated simulations for a
number of parameter points and using interpolation and extrapolation to cover 
the entire parameter plane. Consequently, very close to the threshold region 
(that is, for $m_Y - m_X - m_t \lesim 5\gev$) our results may not be reliable.}.
\item Moderately large values of $m_Y$, $m_Y \lesim 600\gev$, and small 
values of $m_X$. For these values of $m_Y$, the $Y\bar{Y}$ pair is produced 
mostly at rest and the signal selection becomes difficult for small mass 
differences $m_Y-m_X$, when the top quark from the decay $Y \to t X$ is 
quite soft.
\end{enumerate}

In comparison with Ref.~\cite{Han:2008gy}, we obtain somewhat lower values for 
the significance $S/\sqrt{B}$, as a consequence of having performed a more 
realistic simulation that includes QCD radiation (through parton showering) 
and jet smearing.
These effects make it more difficult to devise clean kinematic selection
variables for the signal and result in more background from the tail of 
smeared distributions. 
We have also explored the mass reconstruction scheme proposed in 
Ref.~\cite{Han:2008gy} and the variable $M_{T2}$~\cite{mass2}. 
We have found them to be useful in certain respects and 
complementary to the combination of our cuts. Further 
optimization would depend on detailed (experimental) simulations, 
which we leave for future studies.

In summary, we have found that, at 14 TeV with an integrated luminosity of 
100 fb$^{-1}$, a scalar top partner can be observed at the 5$\sigma$ level 
(or better) for a mass up to 675$\gev$ if $M_X=100\gev$. 
This translates into 945~GeV for a spin-1/2 top partner.
At 8 TeV with an integrated luminosity of 20 fb$^{-1}$, it is possible to 
achieve a 5$\sigma$ discovery for a scalar top with a mass up to 480$\gev$. 
This corresponds to 660$\gev$ for a spin-1/2 top partner.
These results are summarized in Table~\ref{tab:reach}.

%-----------------------------------------------------------------------------
\begin{table}[t]
\renewcommand{\arraystretch}{1.3}
\centering
\begin{tabular}{|c||c|c|}
\hline
 $\sqrt{s}$ & spin-0 & spin-1/2 \\
\hline
 8 TeV   & $ 480 \gev$ &   $660 \gev$   \\
 14 TeV & $ 675 \gev$  & $945 \gev$   \\
\hline
\end{tabular}
\mycaption{The 5$\sigma$ discovery reach for spin-0 and spin-1/2 
top partners at 8 and 14 TeV with integrated luminosities of 
20 and 100 fb$^{-1}$, respectively. $m_X = 100 \gev$ is assumed.
}
\label{tab:reach}
\end{table}
%-----------------------------------------------------------------------------

%%%%%%%%%%%%%%%%%%%%%%%%%%%%%%%%%%%%%%%%%%%%%%%%%%%%%%%%%%%%%%%%%%%%%%%%%%%%%%

\section{Determination of Model Properties}
\label{props}

\subsection{Masses} 
\label{masses}
The independent determination of the $Y$ and $X$ masses in
processes of the type 
\begin{equation}
pp \to Y\bar{Y} \to f\bar{f}XX,  \label{eq:2p}
\end{equation}
where $f$ is a SM fermion, is a difficult problem because of the lack of
kinematic features for the under-constrained system. Several methods have 
been proposed in the literature \cite{mass,mass2,mass3,mass4,Han:2009ss}, 
either based on global event variables such as 
$M_{\rm eff} = \sum_{i \in \rm vis.} p_{T,i} + \pslash_T$, 
on the variable $M_{\rm T2} = \min_{{\bf p}_{ T,X_1}^{} + {\bf
p}_{T,X_2}^{} = \SLASH{\mbox{\scriptsize\bf p}}{.2}_{\rm T}} \left\{ \max
\,\bigl( M_{\rm T}^{\ell^+,X_1}, M_{\rm T}^{\ell^-,X_2} \bigr) \right\}$
\cite{mass2} or variants thereof, or on likelihood fits to the complete event
information \cite{mass3}. 
It was found that, for $m_Y \sim {\cal O}(300 \gev)$ and a sample of a few 
tens of thousands of signal events at $\sqrt{s}=14\tev$, the mass difference 
$m_Y-m_X$ can be determined to a precision of a few per cent, while the 
absolute mass scale has an uncertainty of roughly 20--30\% \cite{mass3,mass4}.
If the $Y\bar{Y}$ state could arise from the decay of a new resonance of 
known mass, it would help to constrain the kinematics and thus to determine 
the masses of $Y$ and $X$ as well \cite{Han:2009ss}. 
More details can be found in the cited papers.

\subsection{Spin} 

The spin of the $Y$ particle can be probed through the
characteristics of the $Y\bar{Y}$ production process. For instance, the
$Y\bar{Y}$ production cross section strongly depends on the spin \cite{spinx}.
However, unknown model-dependent branching fractions and the mass uncertainty of
order 30\% can lead to ambiguities in the determination of the spin from the
measured total production rate. Instead, one can largely avoid such problems 
by investigating the shape of suitable differential distributions.
In particular, the two variables described below are effective for this purpose.
\begin{itemize}

\item[(1)] Scalar and fermion $Y$ pair production can be distinguished with the
observable
\begin{equation}
\tanh(\Delta y_{t\bar{t}}/2), \qquad \Delta y_{t\bar{t}} = |y_{bjj} - y_{b\ell}|,
\label{eq:dy}
\end{equation}
which is constructed from the rapidities of the visible decay products of the
hadronically decaying and the leptonically decaying top quarks. 
In general, there is a combinatorial ambiguity in identifying the $b$-jets and
light-flavor jets as the decay products from one of the two top quarks. 
Given our event reconstruction scheme discussed in the previous section, one
can resolve this ambiguity by assuming that the hadronically decaying top 
quark is made up from the two light-quark jets and the $b$-jet for which 
$m_{bjj}$ is closest to $m_t$. 
The remaining  $b$-jet and the lepton are then identified as the
decay products of the other top quark.

The variable in \eqref{eq:dy} is
closely related to the proposal by Barr in Ref.~\cite{Barr:2005dz}, $\tanh
(|\eta_f-\eta_{\bar{f}}|/2)$, where $\eta_f$ is the pseudorapidity of the SM
fermion from the decay $Y \to Xf$.
This variable approximately traces the production angle $\theta^*$ 
between one $Y$ and the beam axis in the center-of-mass frame.
In the $q\bar{q} \to Y\bar{Y}$ channel, the $\theta^*$ distribution 
has a clear dependence on the $Y$ spin, as can be seen in the 
formulae
\begin{align}
\frac{d\sigma}{d\cos\theta^*}[q\bar{q} \to Y\bar{Y}] 
&
\propto 1-\cos^2\theta^*, %\\
& \text{for scalar $Y$ (spin 0),} \qquad \\
\frac{d\sigma}{d\cos\theta^*}[q\bar{q} \to Y\bar{Y}]
&
\propto 2 + \beta^2_Y(\cos^2\theta^*-1),
& \text{for fermionic $Y$ (spin $\tfrac{1}{2}$),} \,\, 
\end{align}
where $\beta_Y$ is the velocity of the produced $Y$ particles. The difference
stems from the fact that scalars are produced in a $p$-wave, whereas for
fermions the $s$-wave contribution is dominant.
In contrast to Ref.~\cite{Barr:2005dz}, the definition \eqref{eq:dy} is based 
on the rapidities rather than the pseudorapidities, to account for the fact 
that the produced top quarks are massive.

In the physical process $pp \to Y\bar{Y}$, only a subdominant fraction of the
events originates from $q\bar{q}$ annihilation, but as can be seen in
Fig.~\ref{fig:dy} the effect is still noticeable (compare the lines for model 
i with the other cases).
%-----------------------------------------------------------------------------
\begin{figure}[tb]
\centering
\psfig{figure=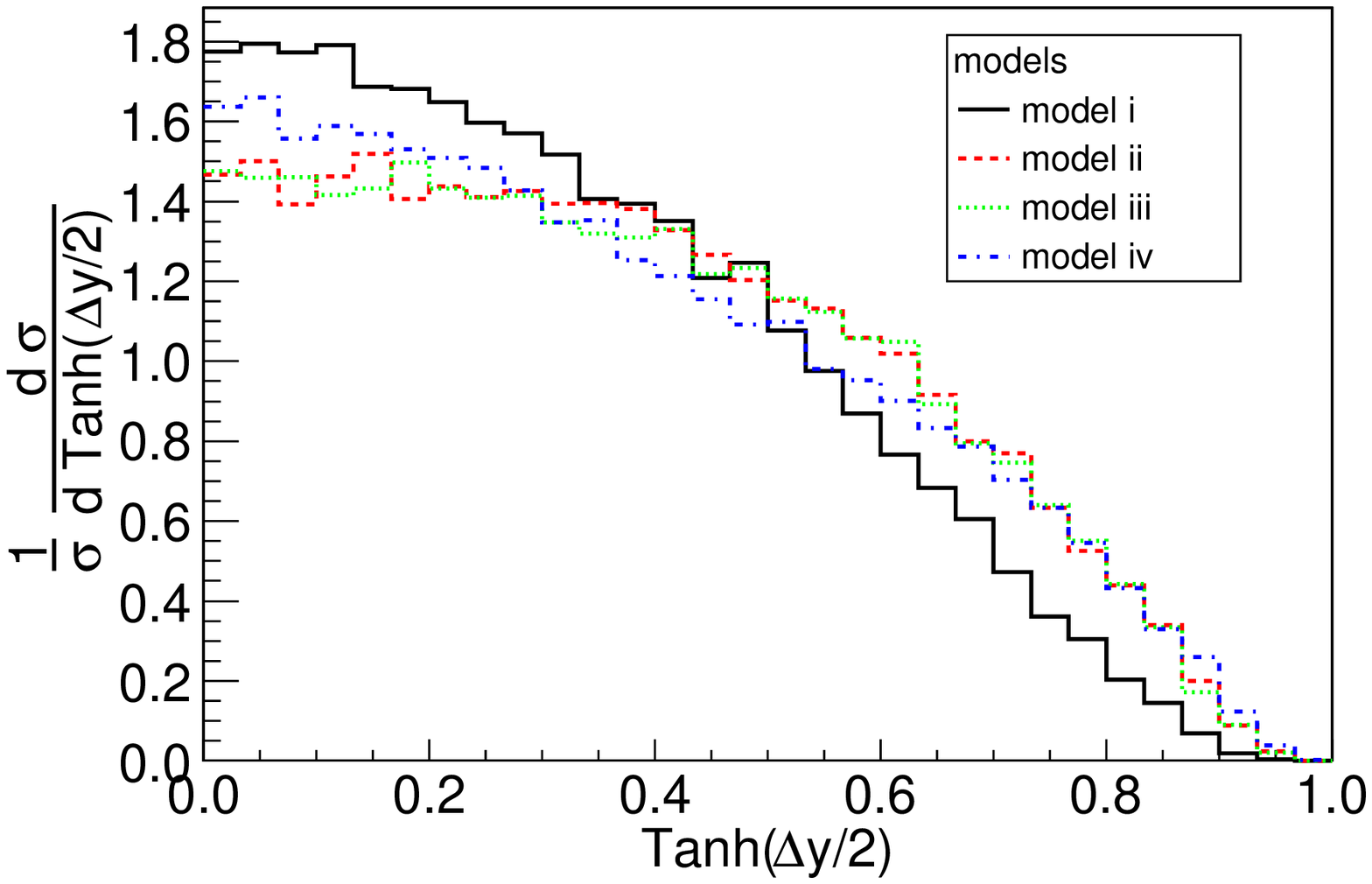, width=13.9cm}
\vspace{-.5em}
\mycaption{Distribution of $\tanh(\Delta y_{t\bar{t}}/2)$ for the different
top-partner scenarios listed in Table~\ref{tab:models1}, for $m_Y=300\gev$,
$m_X=100\gev$, and $\sqrt{s}=14\tev$. 
For comparison, all distributions have been normalized to unity.
\label{fig:dy}}
\end{figure}
%-----------------------------------------------------------------------------

\item[(2)] There is no appreciable difference between fermionic and vector $Y$ pair
production in the $\tanh(\Delta y_{t\bar{t}}/2)$ distribution. However, these
two cases can be disentangled by means of a variable that measures the
effective hard scattering energy \cite{Chen:2010ek}. One such observable is 
the effective mass, a scalar sum over momenta:
\begin{equation}
M_{\rm eff} = \sum_{i \in \rm vis.} p_{T,i} + \pslash_T,
\end{equation}
where the sum runs over all visible objects (jets and leptons in this case).

The usefulness of this variable follows from the fact that the partonic
cross section for the pair production of massive vector particles grows with 
the partonic center-of-mass energy like $\hat{s}/m_Y^4$, whereas for fermions 
it has the usual $1/\hat{s}$ behavior in the high-energy limit. In fact, for 
very large values of $\hat{s}$ the vector $Y\bar{Y}$ production would violate 
the perturbative unitarity limit, a sign that additional new massive 
resonances will appear and modify the production amplitude. The presence of 
such resonances is generally expected in dynamical models such as the one 
proposed in Ref.~\cite{Cai:2008ss}. However, their masses may be beyond the 
reach of the LHC, depending on the value of $\hat{s}$ when the unitarity limit 
is eventually reached. We have estimated this limit for vector $Y$ pair 
production, conservatively assuming $s$-channel dominance in obtaining 
$\hat{s} \gesim (4.8\tev)^2$. Thus, one can assume that the new heavy
resonances have masses of ${\cal O}(5\tev)$, which make their potential
contribution to the process $pp \to Y\bar{Y}$ completely negligible, since 
only a fraction of order $10^{-6}$ of events have partonic center-of-mass 
energy of this size or larger at $\sqrt{s}=14\tev$.

%-----------------------------------------------------------------------------
\begin{figure}[tb]
\centering
\psfig{figure=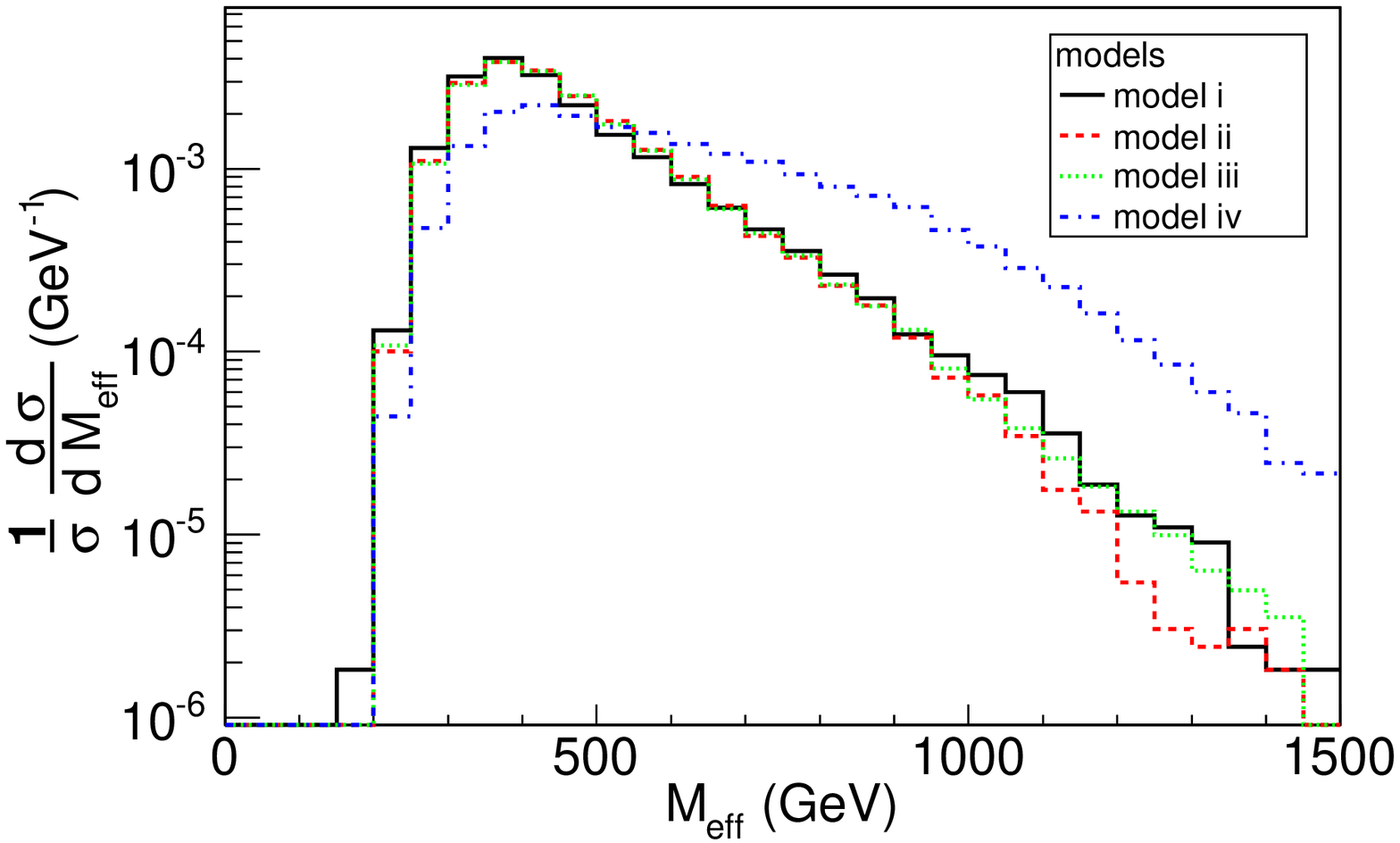, width=13.9cm}
\vspace{-.5em}
\mycaption{Distribution of $M_{\rm eff}$ for the different
top-partner scenarios listed in Table~\ref{tab:models1}, for $m_Y=300\gev$,
$m_X=100\gev$, and $\sqrt{s}=14\tev$. 
For comparison, all distributions have been normalized to unity.
\label{fig:meff}}
\end{figure}
%-----------------------------------------------------------------------------
The $M_{\rm eff}$ distribution is shown in Fig.~\ref{fig:meff} for the 
different model scenarios. A very distinctive difference can be observed 
between the cases of vector and fermionic or scalar top partners.

\end{itemize}

Let us quantify the discriminative power of these variable in an example: 
the process $pp \to Y\bar{Y} \to t\bar{t}XX$ for $m_Y=300\gev$ and
$m_X=100\gev$ at $\sqrt{s}=14\tev$. The simulation and event selection have 
been performed as described in the previous section, with \emph{Cut(1)} in
Eq.~\eqref{eq:cut1}. We have assumed the cross section for scalar $Y$ pair 
production (model i). Note that this scenario is not ruled out by current 
LHC results.
We have not used the total event rate for model discrimination, to avoid
ambiguities due to unknown branching fractions.

We have carried out the discrimination between two different spin assignments 
by computing the $\chi^2$ value for the binned $\tanh(\Delta y_{t\bar{t}}/2)$ 
and $M_{\rm eff}$ distributions, using three bins in both cases. The result 
can be expressed in terms of the integrated luminosity ${\cal L}_{5\sigma}$ 
necessary for achieving a 5$\sigma$ statistical significance:

\vspace{\medskipamount}
\renewcommand{\arraystretch}{1.2}
\begin{tabular}{lll}
14 TeV:
& scalar $Y$ versus fermion $Y$: & ${\cal L}_{5\sigma} = 9.4\text{ fb}^{-1}$, \\
& scalar $Y$ versus vector $Y$: & ${\cal L}_{5\sigma} = 0.8\text{ fb}^{-1}$, \\
& fermion $Y$ versus vector $Y$: & ${\cal L}_{5\sigma} = 0.7\text{ fb}^{-1}$. 
\end{tabular}
\vspace{\medskipamount}

\noindent
For the current 8~TeV run, a 5$\sigma$ discrimination requires the following
integrated luminosities:

\vspace{\medskipamount}
\begin{tabular}{lll}
8 TeV:
& scalar $Y$ versus fermion $Y$: & ${\cal L}_{5\sigma} = 72\text{ fb}^{-1}$, \\
& scalar $Y$ versus vector $Y$: & ${\cal L}_{5\sigma} = 8.1\text{ fb}^{-1}$, \\
& fermion $Y$ versus vector $Y$: & ${\cal L}_{5\sigma} = 5.2\text{ fb}^{-1}$. 
\end{tabular}
\vspace{\medskipamount}

\noindent
The numbers refer to the purely statistical significance. However, at this level
of precision, systematic errors may be important. A potentially large systematic
effect stems from the uncertainty of the new-particle masses, $m_Y$ and $m_X$.
While the mass difference $m_Y-m_X$ can be determined rather precisely,
the overall mass scale can be measured with only 20--30\% accuracy; see
section~\ref{masses}. We have estimated the effect of this uncertainty by
comparing two event samples with $(m_Y,m_X)=(300,100)\gev$ and 
$(m_Y,m_X)=(400,200)\gev$, which differ in $m_Y$ by roughly 30\%. 
We have found that this mass uncertainty reduces the statistical significance
of the spin discrimination by about 20\%;  
the values of ${\cal L}_{5\sigma}$ that account for this systematic error
are about 50\% greater than those quoted above. 

In conclusion, the determination of the spin of the top partner, $Y$, is 
possible with very moderate amounts of data.
On the other hand, the distinction between models ii and iii, which both have
a fermionic $Y$ but differ in the spin of the singlet $X$, is
much more difficult. After surveying more than a dozen different kinematic
variables based on the top-quark momenta, we found no significant difference 
between scenarios ii and iii for any of them. This finding agrees with the 
results of Ref.~\cite{Chen:2010ek}.

However, more information can be obtained from observables that are sensitive 
to the top-quark polarization, as will be discussed next.

\subsection{\boldmath $XY$ Couplings}

The chirality structure of the decay $Y \to tX$ (that is, the relative
contributions of left- and right-handed chiral couplings) leaves an imprint 
on the polarization of the top quark, which can be analyzed through angular
distributions of the top-quark decay products. 
This method is particularly effective when the mass difference 
between $Y$ and $X$ is large ($m_{Y} - m_{x} \gg m_{t}$), so that the top 
quark is energetic and therefore the helicity is preserved, reflecting the 
chirality.
For instance, one can look at the angle $\theta'_b$ ($\theta'_\ell$) 
of the $b$ quark (lepton) with respect to the top-quark boost direction
in the top rest frame. Because the $b$ quark is always left-handed, it is 
emitted predominantly in the forward direction ($\cos\theta'_b>0$) if the 
top quark is left-handed, but mostly in the backward direction
($\cos\theta'_b<0$) if the top quark is right-handed. 

In practice, even if the top quark is produced fully polarized in the decay $Y
\to tX$, some of the polarization is washed out by the mass of the top, but 
the $\cos\theta'_b<0$ distribution will still exhibit a characteristic
difference between left- and right-handed $XYt$ couplings.

\begin{figure}[tb]
\psfig{figure=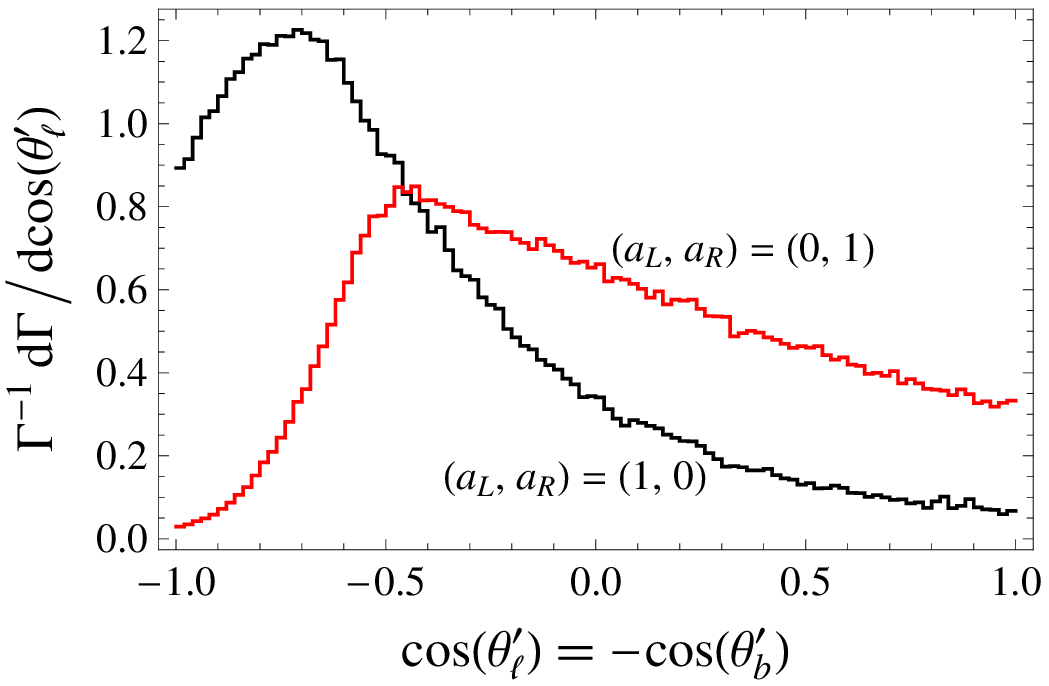, width=8.5cm, bb=0 38 300 196, clip=true}%
\rput[l](-2,4){model i}%
\psfig{figure=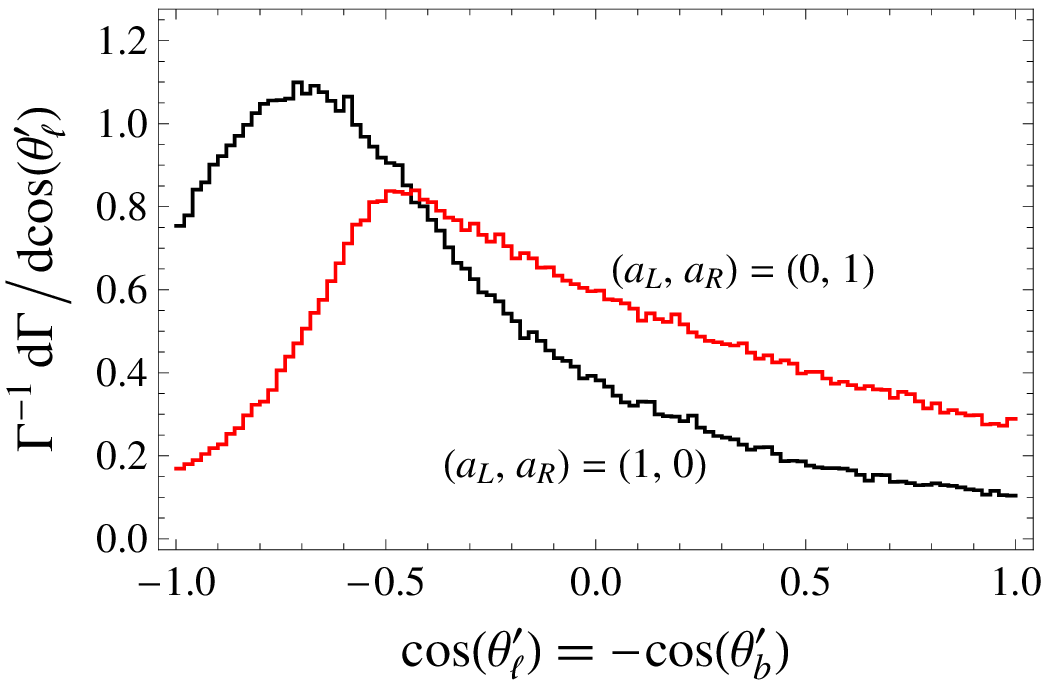, width=7.9cm, bb=21 38 300 196, clip=true}%
\rput[l](-2,4){model ii}\\
\psfig{figure=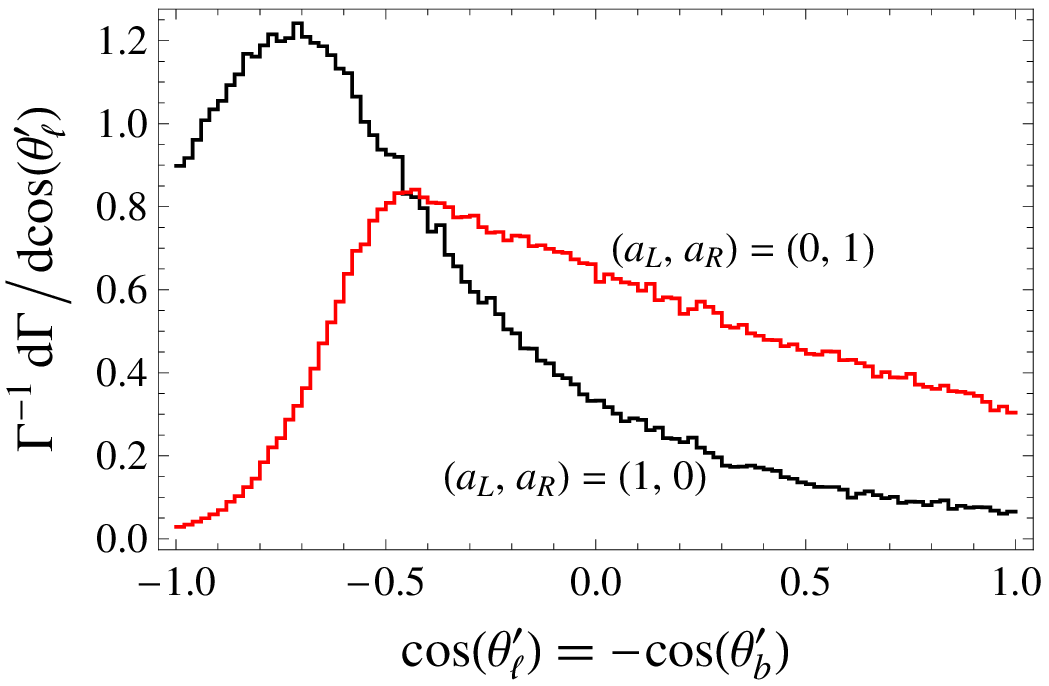, width=8.5cm, bb=0 0 300 196}%
\rput[l](-2,5){model iii}%
\psfig{figure=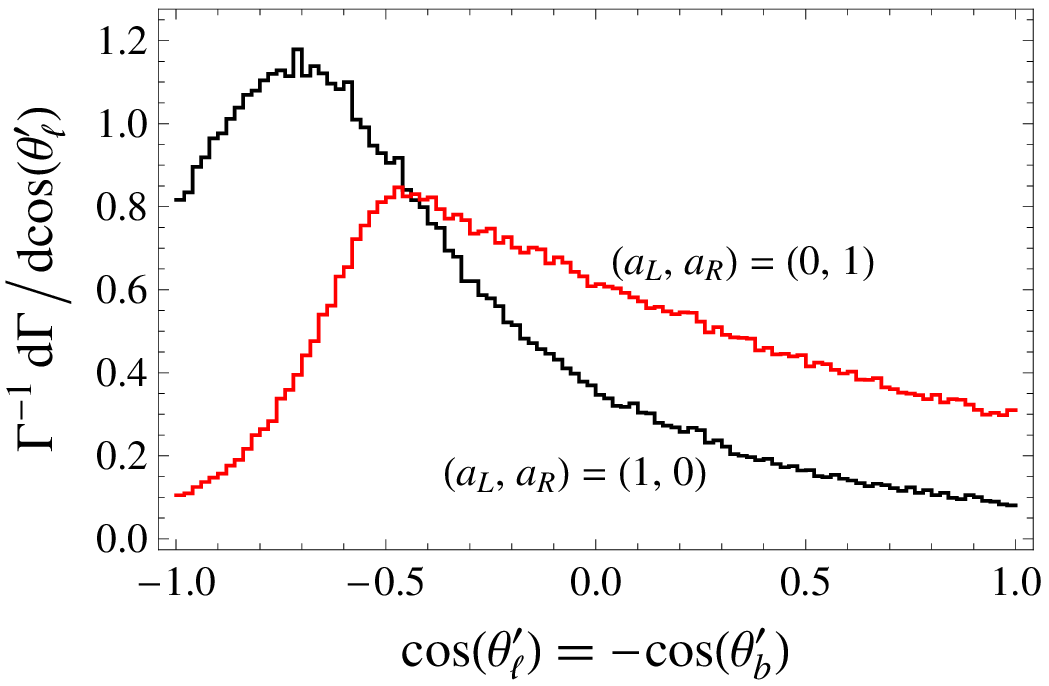, width=7.9cm, bb=21 0 300 196, clip=true}%
\rput[l](-2,5){model iv}%
\vspace{-.5em}
\mycaption{Parton-level angular distribution of the top-quark decay products 
in the $b\ell$ rest frame, for the decay chain $Y \to X t \to X b \ell^+ \nu$.
The four panels show the results for the four scenarios in
Table~\ref{tab:models1}, for the two coupling choices $a_L=1,\,a_R=0$ (black) 
and $a_L=0,\,a_R=1$ (red). The input mass parameters are 
 $m_Y=400\gev$ and $m_X=10\gev$. 
The distributions have been normalized to unity.
\label{fig:pol}}
\end{figure}

In the following, we shall illustrate this behavior using a parton-level 
simulation with
{\sc CalcHEP}\footnote{A more realistic simulation at the level of the
previous section, including parton showering and signal selection cuts, would
require the modification of {\sc Pythia} to include top-quark spin correlation
effects, which we have not attempted to carry out.}. We shall focus on the
leptonically decaying top quark, since it has a cleaner final state. 
The top-quark rest frame cannot be reconstructed because of the unobserved 
neutrino momentum, so we analyze the angular distribution in the rest frame
of the visible $b\ell$ system instead. The results are shown in 
Fig.~\ref{fig:pol}.
%-----------------------------------------------------------------------------
%-----------------------------------------------------------------------------

As Fig.~\ref{fig:pol} shows, the distribution is skewed to smaller values of 
$\cos\theta'_\ell$ or, equivalently, larger values of $\cos\theta'_b$ 
in the case of a left-handed $XYt$ coupling (black curves) than in the 
right-handed case (red curves). 
For a mixed case with non-zero left- and right-handed components, one
obtains a distribution that lies between the black and red curves.
This qualitative behavior is the same for all four spin combinations in
Table~\ref{tab:models1}, although they differ from each other in the
detailed shape of the distribution. In particular, cases ii and iii have 
distinctly different shapes; hence, the analysis of this observable may allow 
one to determine not only the chirality of the $XYt$ coupling but also the 
spin of the $X$ particle.  Such a determination is not possible with 
observables that treat the top quarks as basic objects.

Furthermore, one can probe the chirality even with limited statistics
by using two bins and forming the asymmetry
\begin{equation}
A(x) = \frac{\sigma(\cos\theta'_\ell > x) - \sigma(\cos\theta'_\ell < x)}
        {\sigma(\cos\theta'_\ell > x) + \sigma(\cos\theta'_\ell < x)} \,.
\end{equation}
From Fig.~\ref{fig:pol}, one can see that when $x$ is about $-0.5$
$A(x)$ will be most sensitive to the chirality of the coupling. 
Table~\ref{tab:topasym} shows the asymmetry $A(-0.5)$ for
models i--iv with two choices of the masses $m_Y$ and $m_X$.
The usefulness of $A(-0.5)$ for the determination of the coupling is enhanced
by its relative insensitivity to the spin and mass combinations.

%-----------------------------------------------------------------------------
\begin{table}[hbt]
\renewcommand{\arraystretch}{1.3}
\centering
\newcolumntype{C}{>{\centering\arraybackslash}X}
\begin{tabularx}{.85\textwidth}{|c|CCCC|CCCC|}
\cline{2-9}
   \multicolumn{1}{c|}{}
 & \multicolumn{4}{c|}{$m_Y=400$ GeV, $m_X = 10$ GeV} 
 & \multicolumn{4}{c|}{$m_Y=300$ GeV, $m_X = 100$ GeV} \\
\hline
 & \multicolumn{4}{c|}{Model} & \multicolumn{4}{c|}{Model} 
  \\[-3pt]
$a_L$, $a_R$ & i & ii & iii & iv & i & ii & iii & iv \\
\hline
1, 0 & $-$0.10 & 0.02 & $-$0.10 & $-$0.03 & 0 & 0.15 & 0.04 & 0.10 \\
0, 1 & 0.68 & 0.55 & 0.68 & 0.61 & 0.54 & 0.39 & 0.50 & 0.45 \\
1, 1 & 0.29 & 0.28 & 0.29 & 0.29 & 0.28 & 0.27 & 0.28  & 0.27 \\
\hline
\end{tabularx}
\mycaption{Asymmetry $A(-0.5)$ for models i--iv and two choices 
of the masses $m_Y$ and $m_X$.
}
\label{tab:topasym}
\end{table}
%-----------------------------------------------------------------------------

%%%%%%%%%%%%%%%%%%%%%%%%%%%%%%%%%%%%%%%%%%%%%%%%%%%%%%%%%%%%%%%%%%%%%%%%%%%%%%

\section{Conclusions}
\label{concl}

The SM-like Higgs boson has been observed at the LHC with a relatively light
mass of about 125 GeV. The ``naturalness'' argument of the Higgs sector 
suggests the existence of partners of the SM particles, especially the 
heavy top quark.
The top quark may thus hold the key to new physics associated with the
electroweak symmetry-breaking sector, because of its enhanced coupling to 
the Higgs sector. In this paper, we have systematically categorized the 
generic interactions of a new particle that couples to the top quark and a 
stable neutral particle, which serves as a candidate for cold dark matter. 
We have considered all possible assignments of spin 0, 
$\frac{1}{2}$ and 1 for either of the two new particles.

In the search for new physics involving top quarks and its partners at the LHC,
the experimental signatures  may be distinctive, but challenging to disentangle.
Pair production of the massive top partners leads to a signature of a
$t\bar{t}$ pair plus missing energy, which is difficult to separate from the
large SM $t\bar{t}$ background.  We have presented a set of optimized selection
cuts for isolating this new physics signal at the 8 and 14~TeV runs of the LHC.
We have found that, at 14~TeV with an integrated luminosity of 100 fb$^{-1}$, a
spin-zero top partner can be observed at the 5$\sigma$ level up to a mass of
675~GeV, while for a spin-$\frac{1}{2}$ top partner the reach extends to
945~GeV.

If a process of this type is discovered at the LHC, it will be imperative
to determine the spins and couplings of the new particles, in order to
understand the underlying physics mechanism. We have proposed a strategy to
extract these properties from experimental data by means of suitable
differential distributions of the final-state products. With this approach, 
a spin-0 top partner with mass of about 300~GeV can be discriminated from
spin-$\frac{1}{2}$ and spin-1 particles at the 5$\sigma$ level with a 
luminosity of 10~fb$^{-1}$ at 14~TeV. Furthermore, the structure of the 
coupling that mediates the decay of the top partner into a top quark and a 
massive neutral particle can be analyzed by measurement of the polarization 
of the final-state top quarks.  This method allows one to distinguish clearly 
between left-handed, right-handed, and vector couplings. Most importantly, 
the proposed observables for spin and coupling determination are insensitive 
to unknown branching fractions and depend only mildly on the masses of the 
new particles.

In conclusion, the LHC will allow us to observe and study top partners with 
mass up to about 1~TeV. This program will shed light on the interplay
of the Higgs-boson and top-quark sectors and may elucidate the concept of
naturalness.

%%%%%%%%%%%%%%%%%%%%%%%%%%%%%%%%%%%%%%%%%%%%%%%%%%%%%%%%%%%%%%

\section*{Acknowledgements}

We are grateful to M.~Cacciari and A.~Kardos for providing background
cross sections at 8~TeV. We also thank N.~Christensen for help with 
{\sc CalcHEP} and R.~Mahbubani for correspondence.
This project was supported in part by the National Science Foundation under 
grant PHY-0854782,  
by the US Department of Energy under grant No.~DE-FG02-12ER41832, and by 
PITT PACC.
C.-Y.C was supported in part by the George E. and Majorie S. Pake Fellowship.
Finally, T.H.~is grateful for the hospitality and support of the Aspen Center 
for Physics and the Center for Theoretical Underground Physics and Related Areas (CETUP*) 
during the completion of this work.

%%%%%%%%%%%%%%%%%%%%%%%%%%%%%%%%%%%%%%%%%%%%%%%%%%%%%%%%%%%%%%
%%%%%%%%%%%%%%%%%%%%%%%%%%%%%%%%%%%%%%%%%%%%%%%%%%%%%%%%%%%%%%

\end{document}